# A Novel Thermal Position Sensor Integrated on a Plastic Substrate

A. Petropoulos[1], G. Kaltsas[2,1], D. Goustouridis[1], A.G. Nassiopoulou[1]
[1]Institute of Microelectronics, NCSR-“Demokritos”, Athens, GREECE
[2]Department of Electronics, TEI of Athens, Aegaleo, GREECE

***Abstract*- A thermal position sensor was fabricated and evaluated. The device consists of an array of temperature sensing elements, fabricated entirely on a plastic substrate. A novel fabrication technology was implemented which allows direct integration with read out electronics and communication to the macro-world without the use of wire bonding. The fabricated sensing elements are temperature sensitive Pt resistors with an average TCR of 0.0024/C. The device realizes the detection of the position and the motion of a heating source by monitoring the resistance variation of the thermistor array. The application field of such a cost-effective position sensor is considered quite extensive.**

## I. INTRODUCTION

Modern sensors employ heat transfer in various ways as the way to measure a large number of quantities [1]. This work presents a novel way to exploit heat transfer and temperature sensing in order to construct a position sensor. Motion detection is usually quantified via capacity, electromagnetic, optical or ultrasonic based devices [2-5]. The principle of operation of the presented thermal motion sensor lies in the location detection of a heat source by monitoring the resistance variation of an array of thermistors. The thermistor array is fabricated entirely on a plastic substrate by the combination of standard MEMS and PCB technologies. Therefore the sensing elements are directly integrated to the interface electronics, eliminating the need for wire bonding and die cutting. Excellent thermal isolation between the sensing elements and the substrate is achieved, which enhances the sensors sensitivity and the corresponding response time. Detection of 1D movement in both static and dynamic conditions is demonstrated in this paper, along with a potential expansion of motion detection in more than one dimension.

## II. THE SENSOR ARRAY

### *A. Fabrication Technology*

Each of the sensing elements of the sensor array is a temperature sensitive resistor utilized by a thin Pt film. The resistor is electrically connected to two distinct copper paths of the underlying PCB, each one terminating at a connecting pad. Therefore the resistor is directly connected to the macro-world, without the presence of any intermediates (wire bonding). Measuring the change in the resistance value of the Pt film allows for the calculation of the ambient temperature.

The fabrication technology is schematically shown in Fig. 1 and described in detail elsewhere [6]. Commencing from a PCB substrate, the whole sensor array is build directly on top of it, using standard microelectronic techniques. The most demanding issue when considering PCB as the initial substrate, on which any kind of additional structures are to be fabricated, is the presence of relatively large hypsometric variations. The necessary planarization is performed via a 15μm thick SU-8 layer. Certain holes are formed on the photoresist layer on top of the copper structures, so as to provide an electrically conducting path between copper and the subsequently deposited metal layer. A 20μm layer of the positive photoresist mA-p is utilized as the sacrificial layer. A 30nm/300nm thick Ti/Pt bilayer is then deposited on top of the PCB. Subsequent lift-off removes the Pt from the entire surface except from the areas that were lithographically pre-defined. This way the Pt resistors are fabricated, being electrically connected to the copper tracks through the holes in the SU-8 layer.

An SEM picture of the final device is shown in Fig. 2 where a fabricated array of three thermistors is shown. The thermistor length and width are 1500μm and 100μm respectively while they are located at a distance of 500μm from each other (midpoint-to-midpoint distance). They reside entirely on top of the SU-8 layer. The whole structure presents excellent mechanical endurance, as SU-8 attaches very strongly to the PCB, while it is also very resistive to etchants. Furthermore, the Ti adhesion layer provides enhanced stability to the deposited Pt.

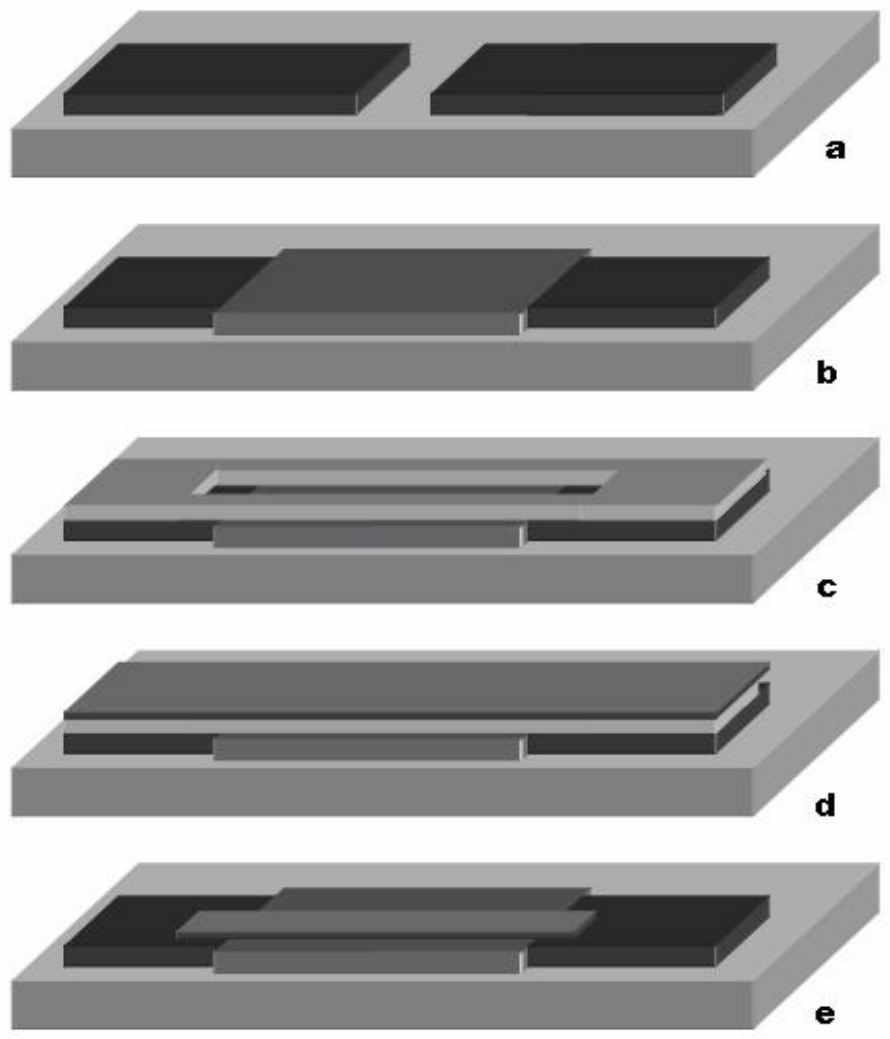

Fig. 1. The fabrication process steps:
**a)** Patterned PCB
**b)** SU-8 planarization layer
**c)** mA-p layer
**d)** Pt deposition
**e)** Lift off

*B. Temperature characterization of the thermistors*

Each thermistor of the fabricated array serves as a temperature sensing element, with a resistance change being a function of the ambient temperature according to the following formula [7]:

$$\frac{R(T)}{R(T_o)} = 1 + \alpha (T - T_o) \quad (1)$$

with $R(T)$ and $R(T_o)$ being the resistance values at a varying temperature $T$ and a reference temperature $T_o$ respectively; $\alpha$ is the thermal coefficient of resistance.

A wide range of measurements has been performed in order to evaluate the steady state behavior of the fabricated sensor to ambient temperature variations. An average TCR of $0.0024/^oC$ was extracted, which is close to relevant TCRs observed for thin Pt films [8]. The characteristic I-Vs of the thermistors reveal a linear relationship between the observed resistance value and the applied power. By combining the aforementioned data, the relationship between the applied power and the developed temperature can be extracted. Therefore the resistance variation of a Pt resistor as a function of both the ambient temperature and the input power can be plotted as shown in Fig.3. The results achieved so far verify that the employed fabrication technology allows for the construction of various thermal sensing devices, which can be directly integrated to the corresponding readout electronics.

## III. Principle Of Operation

*A. Motion definition*

The fabricated thermistor array is employed in order to perform motion detection measurements. This is the first attempt known to the authors to adopt heat transfer as the means to detect displacement. The operating principle of a thermal motion sensor lies in the motion detection of a heating source by recording the induced variation in the resistance values of the thermistors in the array. The heating source moves at a fixed plane parallel to the thermistor array, with the motion axis perpendicular to the thermistors' axes. An elevated temperature distribution is developed in the vicinity of the heating source, so that when a thermistor enters this area the elevated temperature causes the thermistor resistance to increase. The resistance value is directly related to the distance between the thermistor and the heating source therefore the peak value corresponds to the heating source being directly on top of the specific thermistor. It is apparent that the temperature field in the heater vicinity should be highly localized for the whole concept to be feasible.

*B. Heat transfer*

The sensor operation is based upon the convective heat exchange between a heating source and the thermistor array. The contribution of radiative heat transfer to the development of the temperature field is minor, as inferred by simple calculations. The exact shape of the formed temperature field is defined by the properties of both the surrounding medium (air), as well as the sensor material.

A major asset of the fabricated structures is the fact they are composed of low thermal conductivity materials. It is a definitive goal for all temperature sensing applications to minimize heat loss to the substrate as it has a negative impact on the sensor sensitivity and response time. Typically Si-based MEMS devices suffer from a large amount of heat dissipation throughout their structure, due to the presence of the high thermal conductivity silicon substrate. In order to overcome this drawback various ways to achieve thermal isolation to the substrate have been reported, such as the use of vacuum cavities, porous silicon or suspended structures [9-11] but these are most often costly methods employing demanding process steps, while resulting to a high degree of surface anomaly which could impose limitations to further structure fabrication. In the present case, both the SU-8 and the FR4 materials present very low thermal conductivity values of the order of 0.2W/m·K this way enhancing the lateral sensitivity of the device. The medium through which heat transfer takes place is air, which exhibits some very advantageous properties. Its very low thermal conductivity (0.025 W/m·K) leads to a narrow distribution of the temperature in the vicinity of the heating source and prohibits the spread of the thermal energy within the medium which would have a severe negative impact on the device lateral sensitivity. The medium's properties are also the dominating factor regarding the response time of the device. The quantity of interest in this case is the *thermal diffusivity* α defined as:

$$\alpha = \frac{k}{\rho C_P} \qquad (2)$$

where $k$ is the thermal conductivity and $\rho C_p$ is the volumetric heat capacity. Air presents a very high thermal diffusivity value (1938 $10^{-8}$ $m^2/s$), which is far greater than most other fluids allowing for a very low response time of the thermistor signal to the source movement. Furthermore the high thermal diffusivity value inhibits natural convection which is a heat transfer mechanism with severe implications to the observed lateral sensitivity.

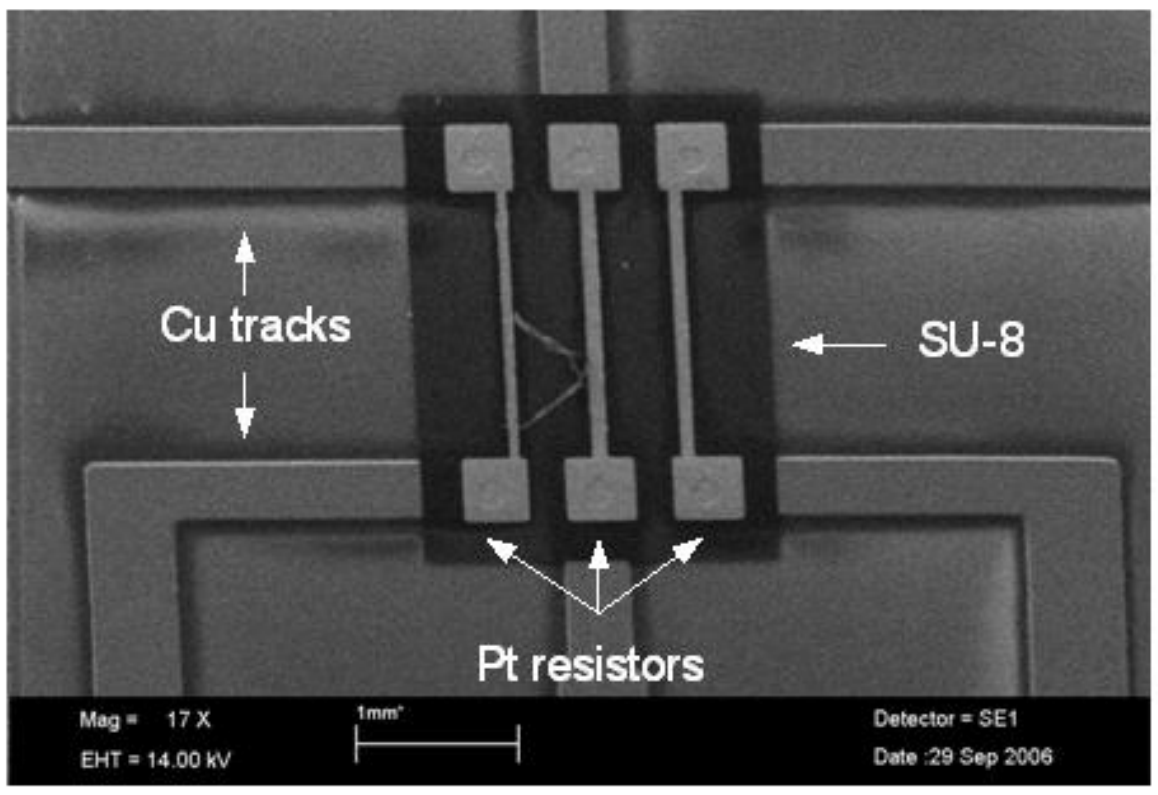

Fig. 2: SEM of the final device. Platinum resistors are in contact with the underlying copper tracks through holes in the SU-8 layer (the latter appears as a dark rectangle)

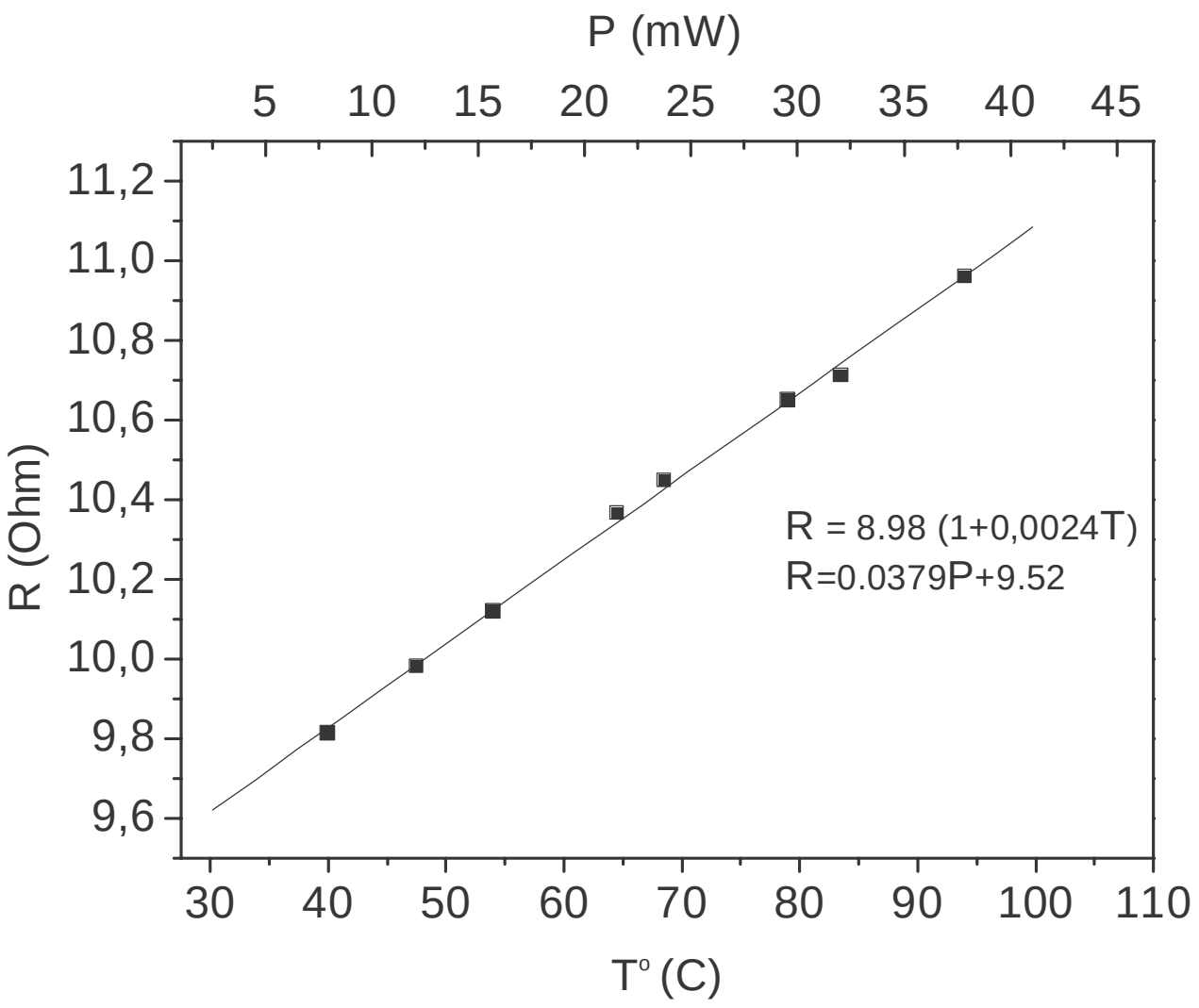

Fig. 3. Resistance change as a function of ambient temperature. The corresponding power is also shown in the graph.

## IV. EXPERIMENTAL DATA

### A. *Static positioning*

The static measurements refer to a heating source of 100μm width performing an 1D movement on top of the sensor array. The plane of motion is parallel to the one defined by the thermistor array at a constant distance of 100μm to 300μm. In Fig. 4 the real time signal obtained at a specific resistor for certain displacement values is shown. The initial increase in the measured resistance is a result of the heating source approaching the sensing element. The signal peak at the distance of 800μm from the commence of the motion indicates the presence of the heating source directly on top of the thermistor; subsequently as the source moves away from it the resistance value decreases.

In Fig. 5 the normalized resistance change $\Delta R/R_o$ of three neighboring resistors (of 500μm distance to each other) with respect to the heating source movement is illustrated. The factor ΔR represents the resistance change and $R_o$ the resistance value at room temperature. A heating source of 100μm width moves in a plane of 1mm fixed distance to the array surface, while the thermal energy produced is 175mW (Fig. 5a). From Fig. 5b we can observe that the maximum signal of the three thermistors R1, R2 and R3 was obtained at 0.4mm, 0.9mm and 1.4mm respectively (distance from the motion's starting point), which indicates the presence of the heating source at the specific points, as expected. The source position at intermediate locations can be extracted by the signal of the adjacent thermistors. The sensors sensitivity at a specific point is defined by the signal of the thermistor that exhibits the highest gradient at the corresponding displacement. A lateral sensitivity value of ±50μm is achieved, which is expected to be significantly improved (down to a few tenths of μm) following the optimization of the device and the corresponding control electronics.

### B. *Dynamic Measurements*

In order to evaluate the dynamic behavior of the sensor, an appropriate measuring setup was designed and fabricated as shown in Fig. 6. The heating source is mounted onto a platform which performs a motion defined by two identical low frequency loudspeakers. The two opposite-facing woofers receive the normal and the inverted signal produced by a waveform generator, so that a predefined one-dimensional movement of the platform is obtained. Accurate control of the vibration over a wide frequency range is maintained. The vertical distance between the heating source and the fixed thermistor array positioned beneath it can be adjusted to less than 1mm. The plane of motion is parallel to the thermistor array, while the axis of motion is perpendicular to the thermistor axes.

Fig. 7 shows the resistance variation of a specific thermistor as a function of time, for a sinusoidal motion of the heating source for various frequencies. As there is no proper way to accurately align the thermistor and the heating source at the initiation of the motion, there is a certain divergence of the sensor signal compared to the sinusoidal motion. The signal peak corresponds to the heating source being directly on top of the specific thermistor, while the minima occur when the source is at the points furthest from the thermistor. Within each period of the source movement, it passes twice over the same spot, therefore the period of the measured signal is half that of the motion. It is apparent that the higher the frequency of motion the lower the resistance variation. The maximum detected signals were of frequencies up to 10 Hz. This clearly depicts the fast response of the device, which is due to the low heat capacity and the low thermal conductivity of the device material as well as to the high thermal diffusivity value of air

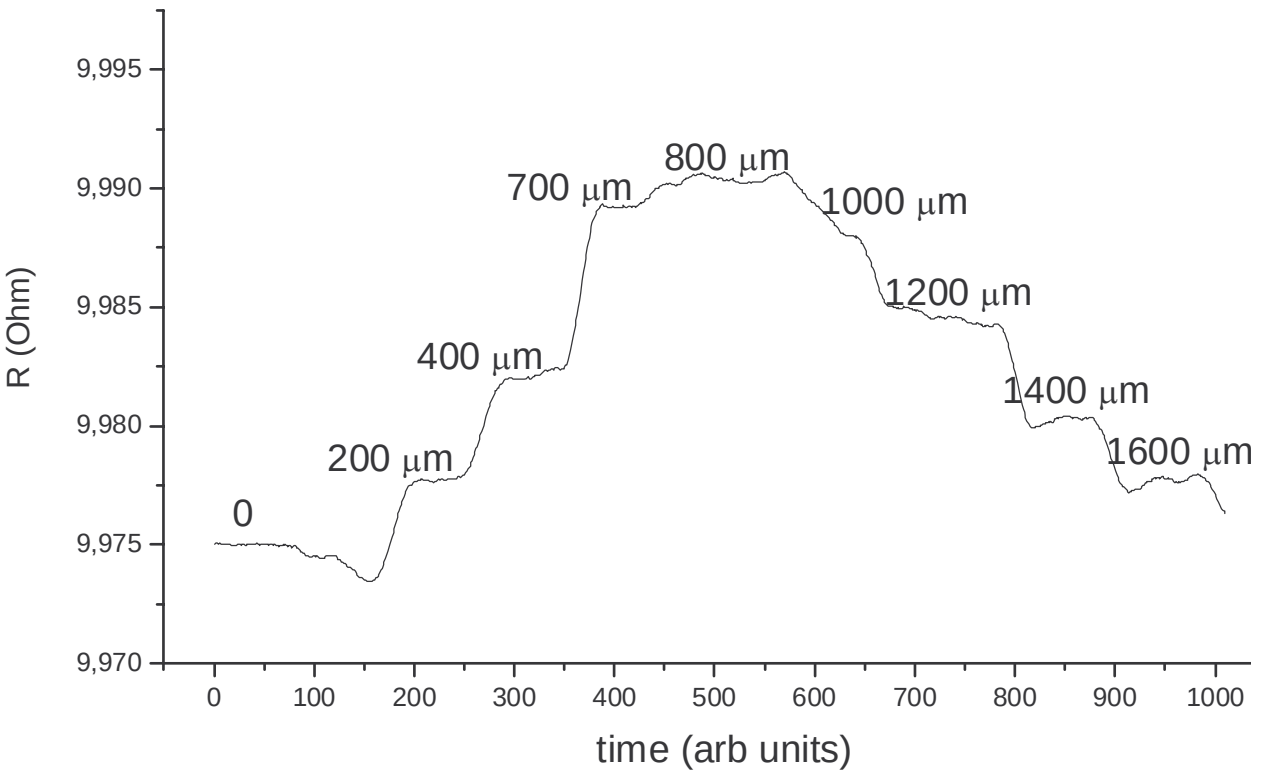

Fig 4. The resistance value of a single thermistor, as the heating source approaches in successive steps of 200μm and 300μm as indicated in the graph.

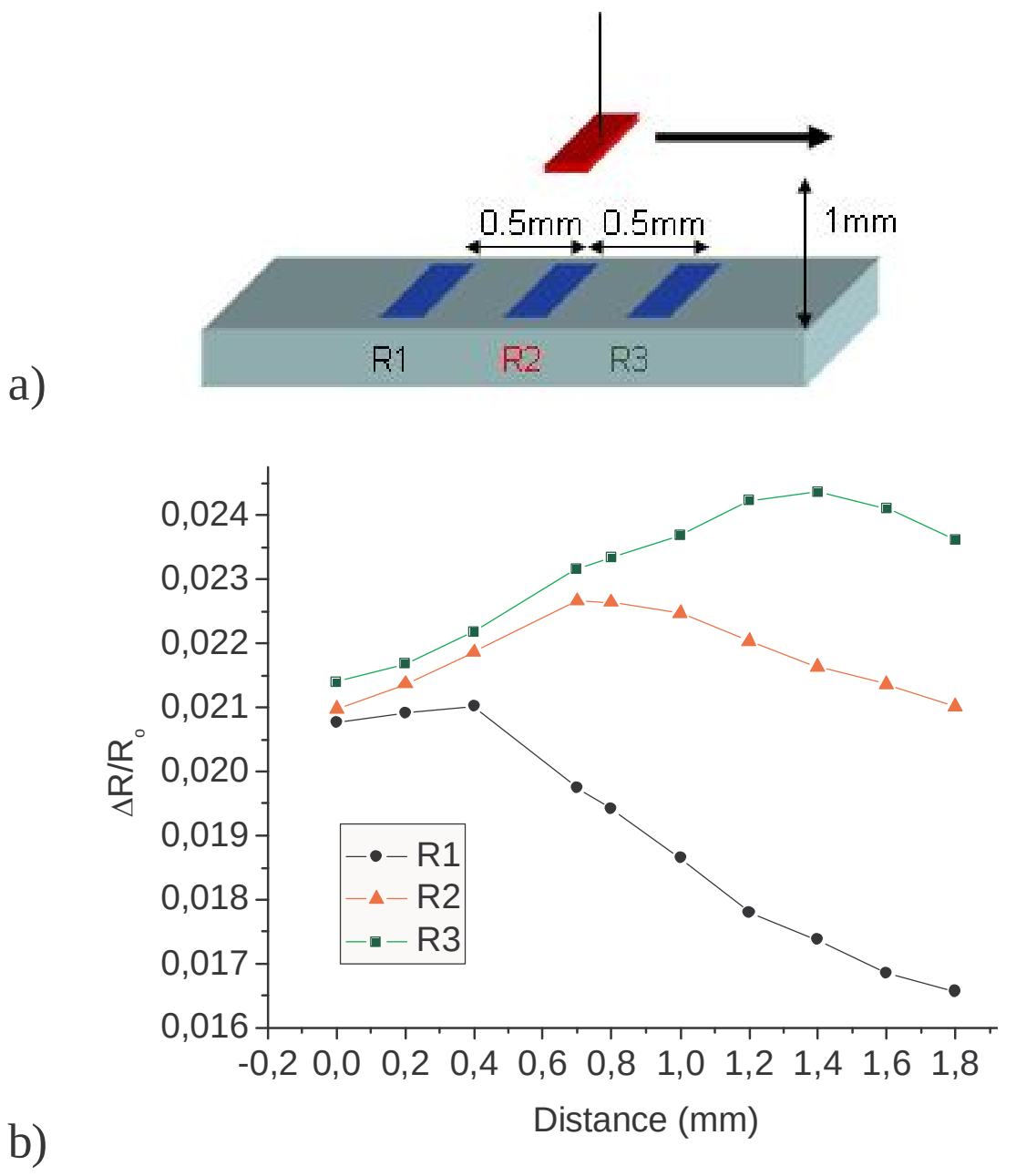

Fig 5. a) Schematic of the measurement setup. The thermistor array remains fixed, while a heated source of 175mW power is mounted on the top plate. The movement of the source is indicated by the arrow b) The recorded normalized resistance change $\Delta R/R_o$ with respect to distance for three thermistors, located at a distance of 500μm to each other.

## V. Overall Characteristics and Further Development

Employing heat transfer in order to determine position as presented in this work is indeed a novel approach, however further optimization of the final device is necessary. Although the observed lateral sensitivity is inferior to some well established methods such as laser interferometry [12], several advantages are obtain through the proposed device such as significantly lower cost, as it is quite simple to fabricate and the absence of special material or equipment. Furthermore, the measuring range can be considered relatively large (~60mm) and can be extended much further in the order of several cm with the addition of extra sensing resistors.

The abovementioned analysis regarding the assessment of displacement by exploiting the temperature distribution can be expanded to two dimensions, by the fabrication of an analogous two dimensional thermistor array. However a more drastic modification of the measuring technique would in principle allow the sensor to provide proximity information as well. In this case, the heating source moves vertically on the z-axis, while the lower plate remains fixed. The resistance values of the thermistor array are dependent on the distance between the array and the heating source. The overall combination of the aforementioned 2D displacement and proximity data leads to a scheme that can in principle provide the means to fully record the three dimensional movement of an object. Experiments towards this direction are in progress.

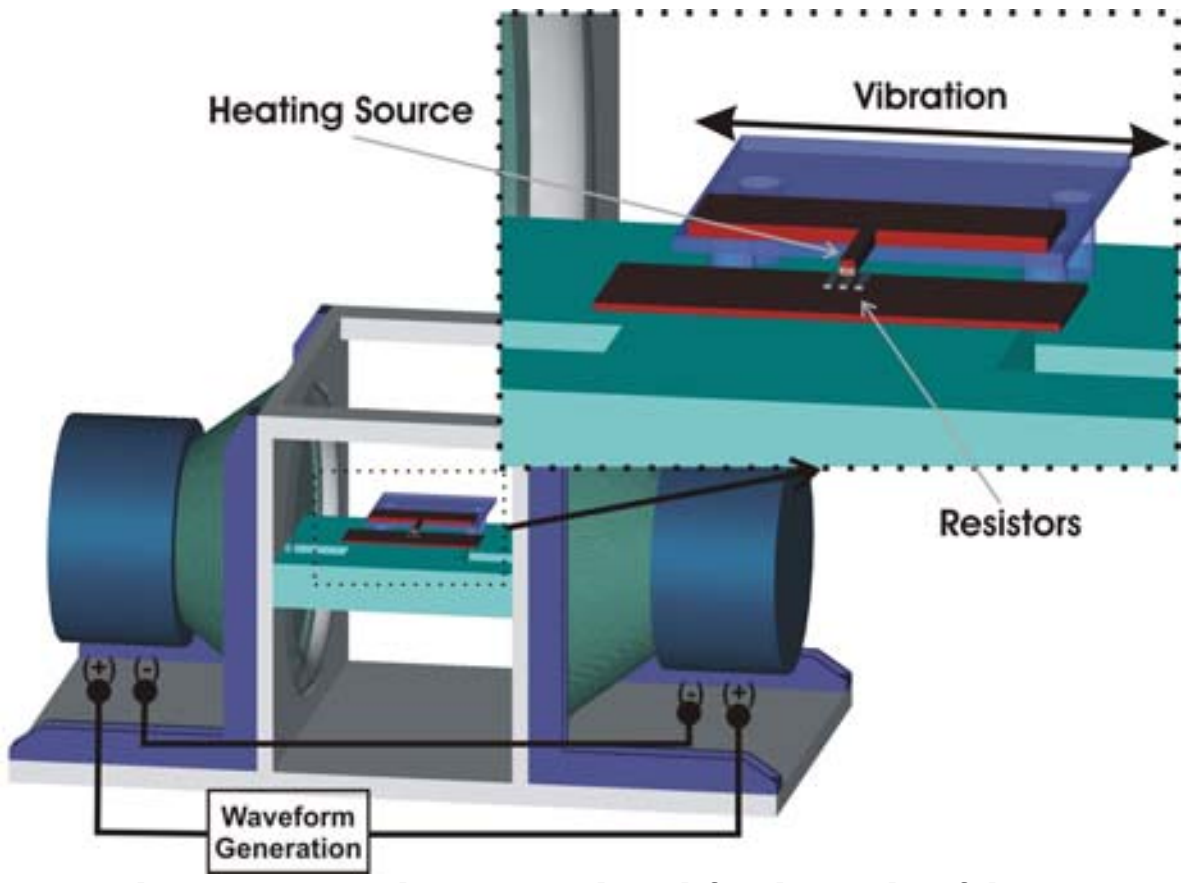

Fig 6. The experimental setup employed for the study of the sensor's dynamic behavior.

## VI. Conclusion

A novel way to exploit heat transfer in order to construct a position sensor is presented and validated. The sensor is based on an array of temperature sensitive Pt thermistors which serve as sensing elements. The fabrication technology allows the thermistors to be directly integrated on the PCB therefore eliminating the need for wire bonding, die cutting and die bonding. The structures comprise of low thermal conductivity materials, a fact that minimizes heat dissipation to the substrate. The thermistor pattern allows for the determination of the one dimensional temperature distribution in a specified area. The motion of a heating source can be inferred by monitoring this temperature distribution. Measurements in both the static and the dynamic mode have been conducted, revealing a lateral sensitivity of ±50μm and a highly satisfactory response time.

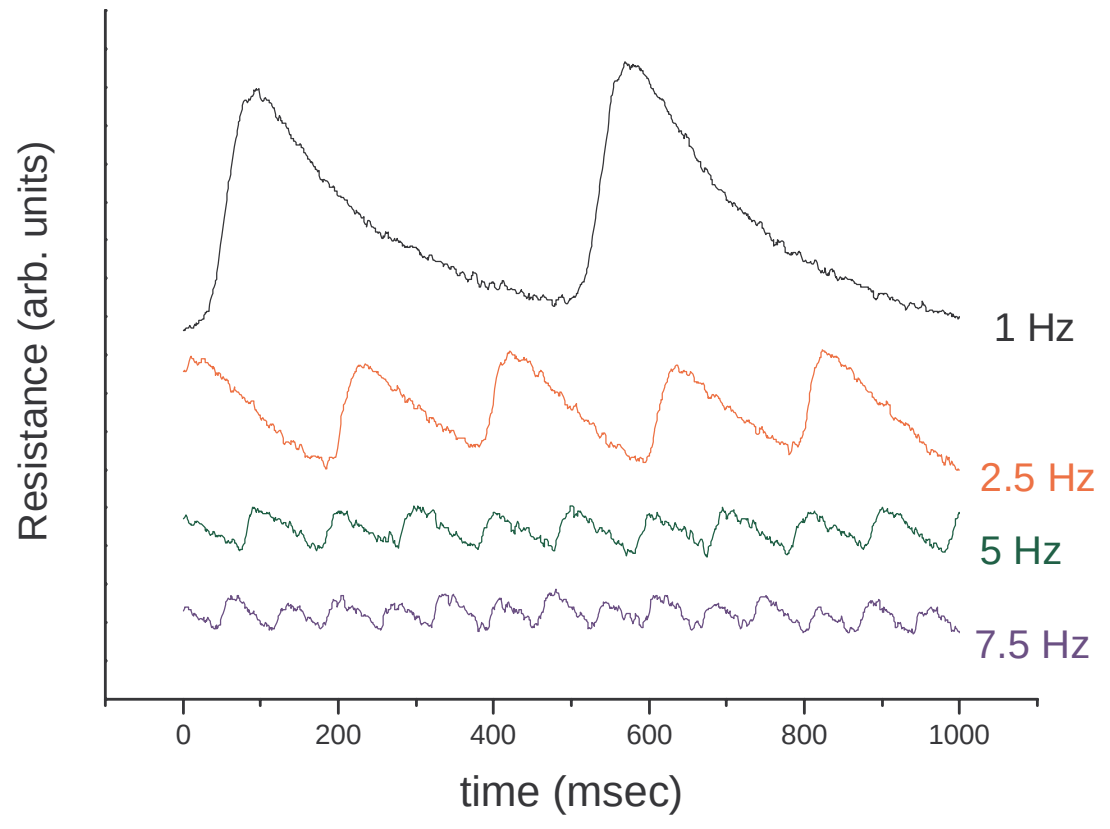

Fig 7. The thermistor resistance while the heating source performs a sinusoidal movement of varying frequency. The actual signals have been shifted for clarity.

A straightforward expansion of the sensor into providing two-dimensional position information could be achieved by an appropriate modification of the thermistor pattern.

## ACKNOWLEDGMENT

This work is co-funded by 75% from the E.E. and 25% from the Greek Government under the framework of the Education and Initial Vocational Training Program-Archimedes and PENED.